# Fourier ptychographic microscopy aided with transport of intensity equation for robust full phase spectrum reconstruction

MIKOŁAJ ROGALSKI,[1,†,*] JUAN MARTINEZ-CARRANZA,[1,†] BARTOSZ GÓRSKI,[1] PIOTR ARCAB,[1] MICHAŁ JÓZWIK,[1] PIOTR ZDAŃKOWSKI,[1] MAGDALENA SOBIEŃ,[2] MARZENA STEFANIUK,[2] SHUN ZHOU,[3,4,5] CHAO ZUO,[3,4,5] AND MACIEJ TRUSIAK[1,**]

[1]*Institute of Micromechanics and Photonics, Warsaw University of Technology, 8 Sw. A. Boboli St., 02-525 Warsaw, Poland*
[2]*Nencki Institute of Experimental Biology BRAINCITY, Pasteura 3, 02-093 Warsaw, Poland*
[3]*Smart Computational Imaging Laboratory (SCILab), School of Electronic and Optical Engineering, Nanjing University of Science and Technology, 210094, Nanjing, Jiangsu Province, China*
[4]*Smart Computational Imaging Research Institute (SCIRI) of Nanjing University of Science and Technology, 210019, Nanjing, Jiangsu Province, China*
[5]*Jiangsu Key Laboratory of Visual Sensing & Intelligent Perception, 210094, Nanjing, Jiangsu Province, China*
[†]*These authors contributed equally.*
*\*mikolaj.rogalski@pw.edu.pl*
*\*\*maciej.trusiak@pw.edu.pl*

**Abstract:** Fourier ptychographic microscopy (FPM) is a pivotal computational imaging technique that achieves phase and amplitude reconstruction with high resolution and wide field of view, using low numerical aperture objectives and LED array illumination. Despite its unique strengths, FPM remains fundamentally limited in retrieving low spatial frequency phase information due to the absence of phase encoding in all on-axis and slightly-off axis (brightfield) illumination angles. To overcome this, we present a novel hybrid approach that combines FPM with the Transport of Intensity Equation (TIE), enabling accurate, full-spectrum phase retrieval without compromising system simplicity. Our method extends standard FPM acquisitions with a single additional on-axis defocused image, from which low-frequency phase components are reconstructed via TIE method, employing large defocus distance to suppress low-frequency artifacts and enhance robustness to intensity noise. High-frequency phase details are recovered through FPM processing, numerically considering LED-specific deviation from plane wave illumination. To additionally compensate for defocus-induced magnification variations caused by spherical wavefront illumination, we employ an affine transform-based correction scheme upon image registration. Notably, by restoring the missing low-frequency content, our hybrid method appears capable of recovering phase values beyond the conventional 0-2π range – an area where conventional FPM techniques often struggle when dealing with optically thick samples. We validated our method using a quantitative phase test target for benchmarking accuracy and biological cheek cells, mouse neurons, and mouse brain tissue slice samples to demonstrate applicability for in vitro bioimaging. Experimental results confirm substantial improvements in phase reconstruction fidelity across spatial frequencies, establishing this hybrid FPM+TIE framework as a practical and high-performance solution for quantitative phase imaging in biomedical and optical metrology applications.

## 1. Introduction

Quantitative phase imaging (QPI) techniques enable label-free, high-contrast visualization of transparent specimens by measuring the optical phase delay introduced by the sample [1–3].

This phase information directly relates to variations in refractive index (RI) and physical thickness, making QPI particularly useful in biological and materials science applications. Similar to conventional optical microscopy, most QPI methods face an inherent trade-off between spatial resolution and field of view (FOV) – achieving high resolution typically requires high numerical aperture (NA) objectives, which inherently limit the imaging area. Conversely, low-NA objectives offer larger FOVs at the cost of reduced resolution. Overcoming this trade-off remains a key challenge in QPI [4].

Fourier ptychographic microscopy (FPM) is a computational QPI technique that addresses this limitation by enabling high-resolution, phase-resolved imaging over a wide FOV [5–8]. FPM is typically implemented using a standard brightfield microscope equipped with a low-NA objective and an LED array for structured, angle-varied illumination [5]. The sample is sequentially illuminated by different LEDs – each corresponding to a unique illumination angle – and the resulting intensity images are iteratively combined in the Fourier domain. This allows reconstruction of the sample's complex optical field with synthetically enhanced resolution, while maintaining the large FOV provided by the low-NA objective.

In recent years, FPM has been rapidly evolving, with advancements focusing on improved reconstruction algorithms [9–12], enhanced illumination schemes [13–15], and more robust calibration procedures [16–18]. However, despite these developments, FPM remains constrained by a fundamental limitation: its inability to accurately retrieve quantitative phase information for low spatial frequency components of the sample. This issue arises because phase information is only encoded in off-axis illumination images. Theoretically, this limitation could be overcome if some illumination angles precisely matched the NA of the objective [15]. However, achieving this condition in low-magnification, LED-array-based FPM systems is challenging because each LED emits a spherical wavefront, leading to spatially varying illumination angles across the large FOV. As a result, this condition can only be effectively satisfied for a small portion of the FOV, further limiting phase retrieval accuracy across the entire imaging area.

To address the challenge of reconstructing low-frequency phase information in FPM, it is possible to integrate the technique with an alternative phase retrieval method. A recent approach combined FPM with digital holographic microscopy [19], but this required substantial modifications to the optical system, including the introduction of coherent object and reference beams. To maintain system simplicity, FPM could instead be combined with a phase imaging technique that is compatible with LED-array-based microscopy. Among available phase retrieval methods, digital phase contrast (DPC) [20,21] and the transport of intensity equation (TIE) [22–24] appear to be the most suitable candidates. However, DPC suffers from the same low-frequency phase retrieval limitations as FPM [25]. In contrast, TIE is well known for its ability to accurately reconstruct low-frequency phase components [24,26], making it a promising candidate for this application.

In TIE formalism, transversal phase variations and longitudinal variations of the intensity are related by a second-order elliptical differential equation (PDE) [27]. Thus, in this PDE, the unknown to solve is the phase. For this, it is necessary to calculate the longitudinal variations of the intensity by calculating the axial derivative of the intensity [23,28]. It has been proved that the axial derivative of intensity can be carried out employing two images acquired at different defocus distances – typically one in-focus image and one out-of-focus image [29,30]. A further advantage of TIE is that the phase reconstruction can be made with coherent or partially coherent illumination [31]. Importantly, since FPM already requires in-focus images, integrating TIE into an FPM system would necessitate acquiring only one additional defocused image.

To the best of our knowledge, a hybrid FPM+TIE approach has not yet been proposed. However, several related studies are worth mentioning. In 2022, S. Zhou et al. [32] introduced a hybrid TIE and Fourier ptychographic tomography method in which two stacks of images (one in-focus and one defocused) were acquired for different illumination angles. TIE was used to reconstruct phase information for each illumination condition, and the resulting complex fields were incorporated into an iterative 3D tomographic reconstruction algorithm with 2D intensity constraints, generating very high computational load and the need for precise angle calibration. In 2023, J. A. Picazo-Bueno et al. [33] proposed a hybrid TIE and synthetic aperture approach, in which images were collected for five different illumination angles (one on-axis and four off-axis) at three defocus levels (in-focus, under-focus, and over-focus). TIE was used to reconstruct phase information for each illumination angle, and the results were combined in the Fourier domain using a synthetic aperture method, slightly increasing the resolution towards FPM like capabilities, yet with significantly less spectral coverage.

Notably, in both studies, TIE reconstruction was performed independently for each illumination angle, effectively limiting the approach to brightfield-only images, as off-axis TIE requires illumination angles smaller than the objective's numerical aperture (NA) [34–36]. Consequently, the achievable resolution in such implementations is inherently lower than that of standard FPM, as only a limited portion of the spatial frequency spectrum is utilized for phase retrieval. Moreover, both methods employed relatively small defocus distances (2 μm and 4 μm, respectively), which ensured accurate reconstruction of high-frequency components but also made the approach susceptible to low-frequency artifacts (LFAs) [30,37,38]. These artifacts primarily result from shot noise in the captured intensity images [39]. A common strategy to minimize the LFAs presence is usually to increase the defocus distance [39]. However, this was not viable in those studies, as it generally leads to a significant degradation in spatial resolution [24] – an unfavorable trade-off for the intended applications.

Another relevant study was conducted in 2024 by J. Wang et al. [40] who proposed a hybrid TIE and DPC method. In this approach, DPC data was supplemented with an additional defocus image for TIE reconstruction. The low-frequency phase information retrieved via TIE was then merged with the high-frequency details from DPC. Interestingly, like FPM, DPC relies on off-axis illumination (commonly implemented with an LED array) and can only reconstruct low-frequency phase components when the illumination NA matches the objective NA [25]. The successful performance of the hybrid DPC+TIE approach suggests that combining FPM with TIE could robustly enhance low-frequency phase recovery.

In this work, we propose a hybrid FPM and TIE method for high-resolution phase retrieval while improving the accuracy of low-frequency phase details. In contrast to previous relevant approaches requiring two [32] or even three [33] angular scanning datasets for brightfield-only illumination, our method utilizes only one standard FPM dataset (i.e., a set of images captured under varying illumination angles with both brightfield and darkfield images) along with only one additional defocused image for TIE reconstruction. Since the TIE step is dedicated solely to recovering low-frequency phase information, the defocus distance can be significantly increased (up to several hundred micrometers), making the method substantially more robust against LFAs. As shown in our results, increasing the defocus distance improves reconstruction stability by reducing the impact of noise, at the cost of reduced TIE resolution. However, this loss is negligible, as the high-frequency details are reliably recovered through FPM. The only added complexity introduced by the increased defocus is the need to correct for magnification differences between the on-axis in-focus and defocused images, caused by the spherical nature of LED illumination. In our implementation, this issue is addressed through a new affine transform-based rescaling procedure.

The performance of the proposed FPM+TIE method is validated through both numerical simulations and experimental measurements, including a phase resolution target, human cheek cells, mouse neurons and mouse brain tissue slice. The results confirm that our method outperforms conventional FPM in retrieving low-frequency phase information. Additionally, it successfully reconstructs phase features with values exceeding the $2\pi$ limit, thereby overcoming one of the fundamental limitations of standard FPM [8,19,41]. Compared to standalone TIE, the proposed FPM+TIE approach achieves an improvement in spatial resolution while substantially minimizing LFAs.

## 2. Methods

Figure 1 illustrates the principle of the proposed FPM+TIE method. The experimental setup (left side of Fig. 1) was implemented using a standard brightfield microscope (Nikon ECLIPSE E200) equipped with a 7×7 LED array (central wavelength: 532 nm, LED pitch: 8.1 mm). The LED array was positioned at a distance of approximately $z \approx 70$ mm below the sample. The sample was imaged using a 4×/0.2NA objective lens, and the images were captured with a Basler a2A5320-23umBAS CMOS camera. The synthetic NA of the system was equal 0.64.

During acquisition, the sample was sequentially illuminated by individual LEDs, resulting in 49 images corresponding to different illumination angles – constituting the standard FPM dataset (Fig. 1(b)). Subsequently, the sample was axially shifted by a defocus distance $\Delta z$, and one additional image was captured under on-axis illumination for TIE reconstruction (Fig. 1(c2)), yielding a total of 50 images per dataset. In this study, we studied the $\Delta z$ distances varying from 20 μm to 300 μm. The defocus was introduced via the used microscope embedded manual stage and controlled with the use of a length gauge (HEIDENHAIN MT 60K, 0.5 μm accuracy).

In conventional FPM reconstruction, the illumination wavefronts are typically assumed to be plane waves incident at various angles. However, this assumption breaks down in the case of TIE when using an uncollimated LED source and capturing images at significantly large defocuses. Under such conditions, the in-focus and defocused images exhibit different geometrical magnifications, which must be corrected for accurate TIE reconstruction. The relative magnification is described by the formula:

$$m = \frac{z + \Delta z}{z} \qquad \text{Eq. (1)}$$

For the largest tested defocus distance, the magnification factor was approximately $m \approx 1.004$. For the measured FOV (3.6 x 2 mm) this could yield up to $\sqrt{3.6^2 + 2^2}/2 \cdot (m-1) = 8.2$ μm transversal displacement between in-focus and defocused images.

To correct for this, the defocused image should be rescaled by a factor of $1/m$. In practice, this correction is custom-implemented by us via an affine transformation, estimated by manually identifying three corresponding reference points in both the in-focus and defocused images. This approach ensured compatibility with the acquired dataset and also compensated for any minor transverse misalignments introduced by imperfect optical alignment or slightly off-axis illumination. In principle, the process of manually estimating an affine transform should be performed only once and then the same correction may be performed for different measurements, providing the good repeatability of the introduced defocus distances.

The FPM+TIE (Fig. 2(e)) reconstruction pipeline begins with TIE phase retrieval (Fig. 2(f)), performed using the in-focus and defocused images acquired under on-axis illumination, employing the FFT solver [42]. This is followed by the standard FPM reconstruction (Fig. 2(d)), carried out using a quasi-Newton optimization algorithm as described in [43,44]. The only modification to the FPM procedure lies in the initialization step: the initial guess (IG) now incorporates the low-frequency phase estimated from the TIE result $\varphi_{TIE}$. Specifically, the IG is defined as:

$$IG(x', y') = \left\lceil \sqrt{I_{cen}(x, y)} \cdot \exp\left(i\varphi_{TIE}(x, y)\right) \right\rceil, \qquad \text{Eq. (2)}$$

where $I_{cen}(x, y)$ is the intensity image captured under central (on-axis) LED illumination and $\lceil \cdot \rceil$ operator denotes resizing of the image to match the final FPM reconstruction dimensions.

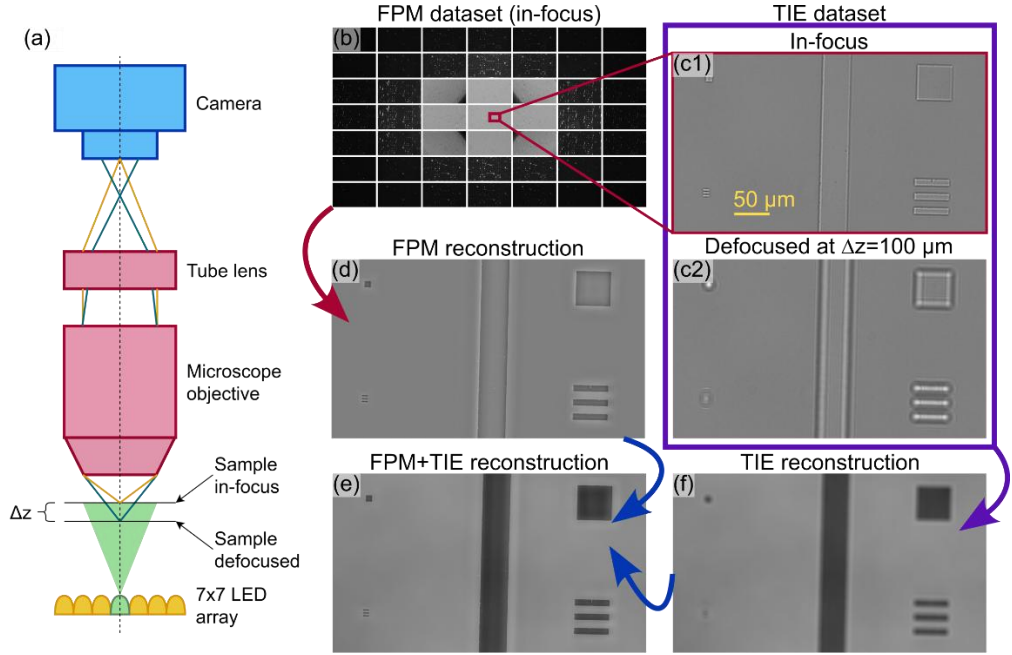

Fig. 1. (a) Scheme of the employed FPM system with images collected in- and out-of-focus. Acquired (b) FPM and (c) TIE dataset along with (d) FPM, (e) FPM+TIE and (f) TIE reconstructions.

## 3. Numerical Simulations

In the FPM technique, images are acquired under varying illumination angles, and each image encodes different components of the sample's spatial frequency spectrum. The set of spatial frequencies captured in a given image is determined by the amplitude and phase transfer functions (ATF and PTF, respectively). According to the formulation provided by J. Sun et al. [15], these transfer functions for brightfield illumination can be expressed as:

$$ATF(u) = \frac{1}{2}[P(u+u_0) + P(u-u_0)] \quad \text{Eq. (3)}$$

$$PTF(u) = i[P(u+u_0) - P(u-u_0)] \quad \text{Eq. (4)}$$

where $u$ denotes spatial frequency coordinates in the Fourier domain, $u_0$ represents the frequency shift associated with the illumination angle, and $P$ is the 2D circular pupil function of the microscope objective.

From Eq. (3), it follows that amplitude information at any frequency $u_n$ can be retrieved by choosing an illumination angle (i.e., $u_0$) such that $u_n$ lies within either $P(u+u_0)$ or $P(u-u_0)$. In contrast, retrieval of low-frequency phase information is more challenging. For small $u_0$ (when $u_0$ is smaller than radius of $P$), the PTF terms $P(u+u_0)$ and $P(u-u_0)$ substantially overlap and cancel each other out (Eq. 4), resulting in negligible PTF response for central frequencies. Therefore, the only condition under which low-frequency phase information can be effectively captured is when the illumination numerical aperture matches the objective NA – referred to as the "matching NA" condition. However, as shown later in the experimental section, this condition is often difficult to fulfill in practical FPM systems.

An alternative approach to encoding low-frequency phase information involves defocusing the sample. For this case, following the derivation by C. Zuo et al. [45], the PTF for coherent, on-axis illumination becomes:

$$PTF(u) = [P(u)\sin(\pi\lambda u \cdot \Delta z)] \quad \text{Eq. (5)}$$

Thus, for any nonzero defocus distance $\Delta z$, the PTF has a nonzero response even at low frequencies, enabling phase retrieval.

Figure 2 demonstrates the effect of the PTF on low-frequency phase reconstruction. A 1024×1024 px size synthetic object composed of parallel phase-only lines of varying widths (256, 64, 16, 4, 2, and 1 pixels) was simulated, as shown in the inset of Fig. 2(a1). An FPM dataset was then generated using a 7×7 LED array, with illumination angles sampled uniformly in the Fourier domain with a step size $\Delta u = 128 \text{ px}^{-1}$. In this model, the central LED (index (4,4)) corresponds to a frequency shift $u_0 = (0,0)$, while LED (3,4) corresponds to $u_0 = (\Delta u, 0)$, and LED (5,6) to $u_0 = (-\Delta u, -2\Delta u)$, and so on. The pupil function radius $R_p$ was defined as:

$$R_P = \sqrt{2}\Delta u \xi \qquad \text{Eq. (6)}$$

where the parameter $\xi$ determines whether the matching NA condition is met (whether $R_p$ matches $u_0$ for any LED). For example, $\xi = 1/\sqrt{2}$ fulfills the matching NA condition for LEDs adjacent to the central one in horizontal and vertical directions; $\xi = 1$ fulfills it for central LED diagonal neighbors; and $\xi = 1.2$ does not fulfill it for any simulated LED.

Figures 2(a)-2(d) present simulated images and their Fourier spectra under various conditions. Fig. 2(a) shows the result for on-axis illumination with $\xi = 1.2$; Fig. 2(b) is for off-axis illumination (LED index (3,5)) also with $\xi = 1.2$. In this configuration, matching NA conditions are not satisfied, and because the sample is purely phase, missing frequencies appear in the spectrum – particularly in the regions where $P(u + u_0)$ and $P(u - u_0)$ overlap, as shown in Fig. 2(b2).

In contrast, Fig. 2(c) shows the case for $\xi = 1$, where the shifted pupil functions do not overlap but still cover low-frequency components. This case may seem counterintuitive as for $\xi = 1$, the NA of the system is smaller than for the $\xi = 1.2$ case, and thus each collected image contains effectively less information. However, since the matching NA conditions are satisfied, the low-frequency phase components are successful retrieved. Fig. 2(d) demonstrates an on-axis illumination case with $\xi = 1.2$ and defocus distance $\Delta z = 100$ μm. Despite the absence of off-axis illumination and the phase-only nature of the object, the image contains retrievable phase information due to the defocus.

Figures 2(e)-2(g) present FPM reconstructions obtained for various $\xi$ values. As shown, the FPM reconstruction accurately recovers low-frequency phase information only for $\xi = 1$, i.e., under matching NA conditions. Even a slight deviation, such as $\xi = 1.01$, leads to significant degradation in low-frequency reconstruction – see missing phase information in the central part of largest lines and the inserts in Figs 2(e)-2(g) showing the central 100x100 px$^{-1}$ part of Fourier domain of reconstruction. Finally, Fig. 2(h) shows the result of the proposed FPM+TIE approach for $\xi = 1.2$ and $\Delta z = 100$ μm, where the missing low-frequency components are successfully restored.

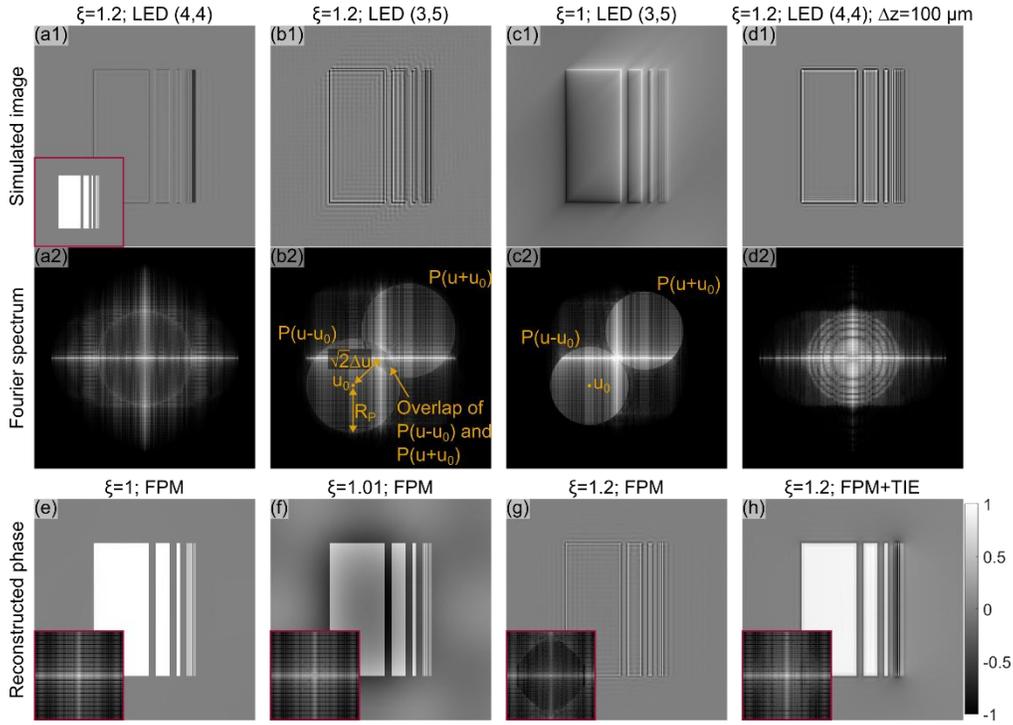

Fig. 2. (a)-(d) Simulated intensity images and their Fourier spectra for (a) on-axis illumination, (b) off-axis illumination when $R_P > \sqrt{2}\Delta u$, (c) off-axis illumination when $R_P = \sqrt{2}\Delta u$ and (d) on-axis illumination and imaging sample in defocus. Insert in (a1) show the simulated phase-only object. (e)-(g) Reconstructed phase with FPM method for ξ equal 1, 1.01 and 1.2 respectively. (h) Reconstructed phase with FPM+TIE method for $\xi = 1.2$. Inserts in (e)-(h) show the central 100x100 px$^{-1}$ part of the reconstructed Fourier spectrum.

To further evaluate the impact of the TIE method on FPM reconstruction, we investigated the accuracy of TIE phase retrieval as a function of defocus distance, as presented in Fig. 3. The simulation was based on the same phase object used in the previous simulation (Fig. 3(a)). TIE reconstructions were performed using images simulated for $\xi = 1.2$ with two different defocus distances: $\Delta z = 20$ μm (Fig. 3(b)) and $\Delta z = 100$ μm (Fig. 3(c)), resulting in higher resolution reconstruction for $\Delta z = 20$ μm case (with 4 px wide lines visible).

Next, we assessed the robustness of TIE under noise. Gaussian noise with standard deviation $\sigma = 0.02$ a.u. was added to the input intensity images. The standard deviations of the corresponding noise-free images were 0.04, 0.11, and 0.13 a.u. for the in-focus, $\Delta z = 20$ μm and $\Delta z = 100$ μm cases, respectively. TIE reconstructions were again performed for both defocus conditions (Figs. 3(d) and 3(e)). For $\Delta z = 20$ μm the presence of noise significantly degraded the reconstruction, with prominent LFAs clearly visible. In contrast, for $\Delta z = 100$ μm, the reconstruction remained robust, showing minimal deviation from the noise-free case.

Fig. 3(f) presents horizontal cross-sections through the reconstructions in Figs. 3(a)–3(e). The plot illustrates that for $\Delta z = 20$ μm case, the higher resolution (with 4 px lines visible) was achieved than for the $\Delta z = 100$ μm case. Additionally, it can be observed that for noise-free $\Delta z = 20$ μm and especially for both $\Delta z = 100$ μm results the negative phase values were obtained around the area of 4 px width lines. This indicates, that even for noise-free conditions, the method may result in inaccurate phase estimation for smaller details.

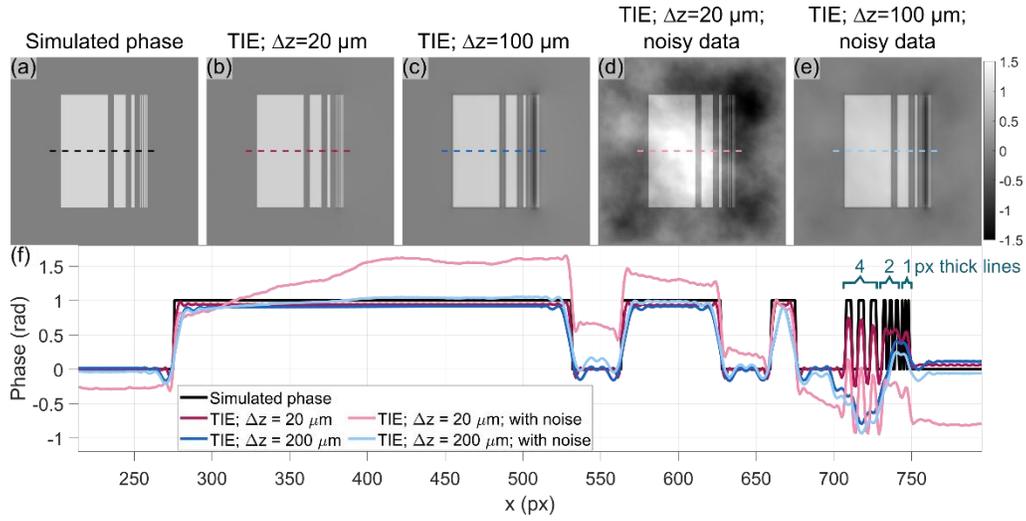

Fig. 3. (a) Simulated phase; (b),(c) TIE reconstructions for noise-free data; (d),(f) TIE reconstructions for noisy data. (f) cross-sections through the (a)-(e) images.

Figure 4 shows FPM-only (Fig. 4(a)) and FPM+TIE reconstructions (Figs. 4(b)–4(e)) based on the TIE results from Fig. 3. Figure 4(f) presents horizontal cross-sections through the corresponding images. As expected, FPM enables recovery of high-frequency information, allowing resolution of the narrowest (1 px wide) lines in the simulated phase object. These details were not resolved in the TIE-only reconstructions due to limited NA (for $\Delta z = 20$ μm or defocus-induced blur (for $\Delta z = 100$ μm).

In the FPM-only result (Fig. 4(a)), low-frequency phase components are missing – particularly in the central regions of the wider lines. When FPM is combined with noisy TIE data at $\Delta z = 20$ μm, the LFAs from Fig. 3(d) propagate into the final result (Fig. 4(d)). In contrast, the FPM+TIE reconstructions for $\Delta z = 100$ μm (Figs. 4(c) and 4(e)) closely match the noise-free result obtained for $\Delta z = 20$ μm (Fig. 4(b)), indicating that higher defocus distances offer increased robustness to noise while preserving phase accuracy for both low- and high-frequency phase information.

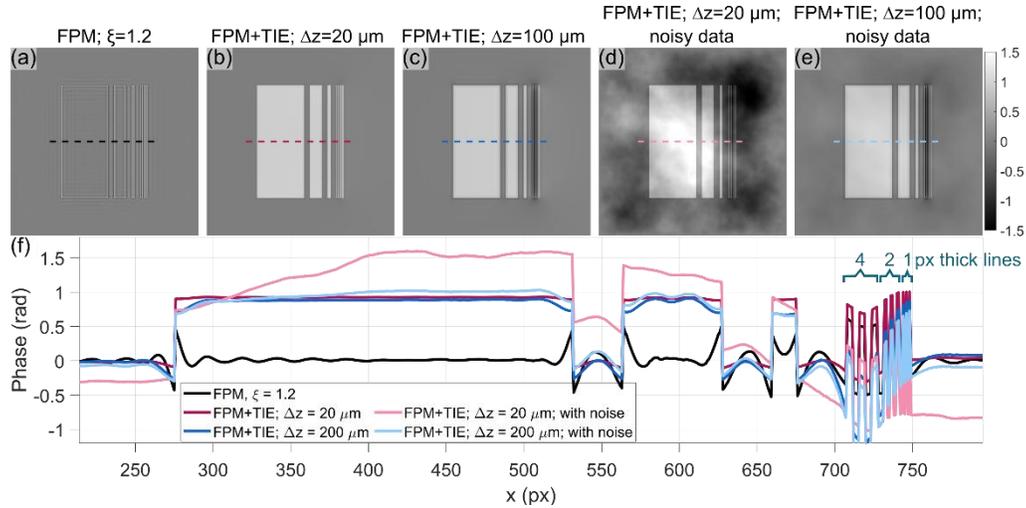

Fig. 4. (a) FPM reconstruction for $\xi = 1.2$; (b),(c) FPM+TIE reconstructions for noise-free data; (d),(f) FPM+TIE reconstructions for noisy data. (f) cross-sections through the (a)-(e) images.

In the experimental part of this study, we utilized defocus distances of 20 μm, 100 μm, and 300 μm to demonstrate the flexibility of the proposed FPM+TIE approach. It is worth emphasizing that the defocus distance used in the TIE component can be easily adjusted to match the specific configuration and requirements of a given imaging setup, including factors such as the objective's numerical aperture, illumination wavelength, sample type, and system noise level. While our simulations provided general insight into the method's robustness across a range of conditions, the optimal defocus distance can be empirically selected in practice to balance reconstruction quality and experimental convenience.

## 4. Experimental Results

Figure 5 shows the experimental validation of the FPM+TIE method for the custom-made phase resolution target (Lyncee Tec, Borofloat 33 glass). Figure 5(a) shows the input intensity image from FPM dataset collected with central LED illumination (LED index (4,4)), where the measured object is of very low contrast with only the edges of elements being visible. Fig. 5(b) also shows the input image from FPM dataset, but this one was collected with diagonal LED (LED index (3,5)), that illuminated the sample with illumination NA close to the objective NA. However, as can be noticed, most of the observed FOV is imaged in brightfield illumination condition, while top-right part of FOV is imaged with darkfield illumination. This results from the fact that the measured sample is not illuminated by the plane wave (as assumed in FPM technique) but with a spherical wavefront coming from a point source which is a singular LED diode. Therefore, each part of FOV is effectively illuminated by the different illumination angles, allowing to achieve the matching NA conditions only for a narrow part of FOV (on the border between brightfield and darkfield illumination area).

Figure 5(c) shows the TIE reconstruction for small $\Delta z = 20$ μm value. As can be observed, the reconstruction was corrupted by strong LFAs, which significantly degrade the obtained phase. Figures 5(d)-5(f) show the TIE, FPM and FPM+TIE results respectively for TIE data acquired with large $\Delta z = 100$ μm. As can be observed, for larger defocus value, the LFAs are significantly reduced compared to smaller defocus TIE. FPM and FPM+TIE methods achieved appreciably higher resolution than large defocus TIE-only reconstruction, with elements from smallest the test group S (2 μm wide) easily resolvable – see zoomed in areas in Figs. 5(e)-(f).

Comparing the low-frequency phase reconstruction, it can be noticed that FPM technique does not recover the phase values in the central parts of the larger test elements, which were successfully retrieved by TIE and FPM+TIE methods. To quantify this, we calculated the average phase value $\pm$ its standard deviation inside three separate square elements of different sizes: large (LE; 250x250 μm), medium (ME; 30x30 μm), and small (SE; 10x10 μm), marked with yellow arrows in Fig. 5(f). The reference height of the elements, measured with a grating interferometer was equal to $0.52 \pm 0.08$ rad.

The obtained absolute phase values, presented in Table 1, show that reconstructed SE and ME phase values closely match the reference ones for FPM+TIE method. The LE phase values are underestimated, which probably was caused due to the LFAs. Nevertheless, the proposed technique still outperforms the FPM-only method, which strongly underestimates larger phase elements, and standalone TIE, which underestimates the smaller phase elements due to resolution limitation.

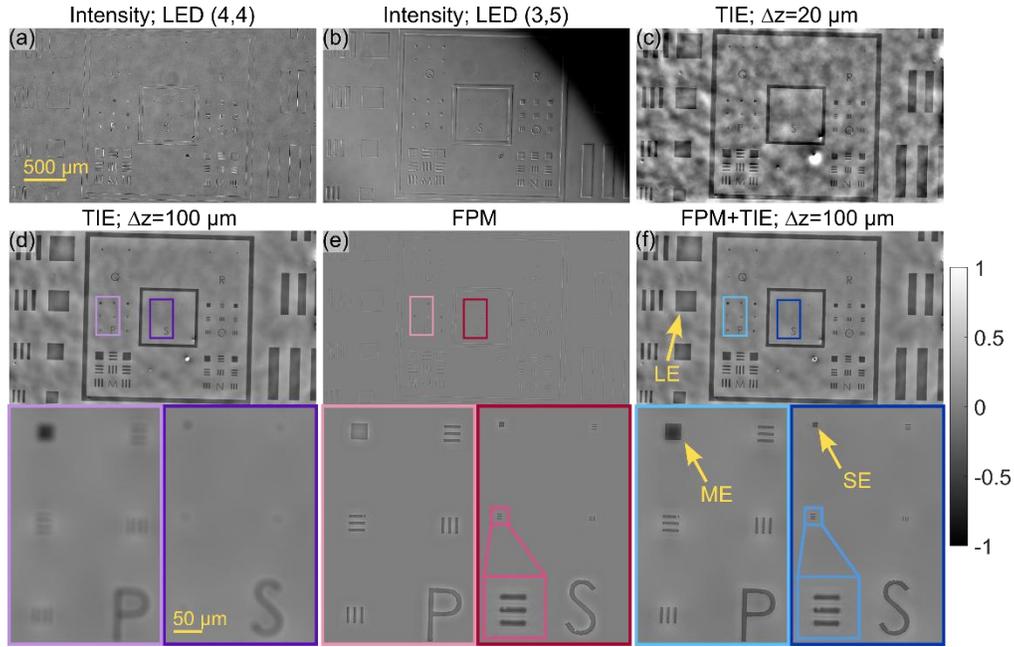

Fig. 5. Collected FPM dataset images with (a) on-axis and (b) off-axis (close to "matching NA" conditions) illumination. (c) TIE phase reconstruction for small Δz with large LFAs visible. (d)-(f) Phase reconstructions with (d) large Δz TIE, (e) FPM and (f) FPM+TIE methods shown in full FOV. Table 1 shows the average phase values for square elements marked with LE, ME, SE in (f).

Table 1. Average absolute phase values $\pm$ their standard deviation for LE, ME and SE square elements shown in Fig. 5(f). The reference phase value of the elements is equal $0.52 \pm 0.08$ rad.

| Method | LE phase [rad] | ME phase [rad] | SE phase [rad] |
|---|---|---|---|
| TIE | $0.22 \pm 0.09$ | $0.54 \pm 0.16$ | $0.32 \pm 0.13$ |
| FPM | $0 \pm 0.02$ | $0.06 \pm 0.06$ | $0.28 \pm 0.05$ |
| FPM+TIE | $0.22 \pm 0.09$ | $0.56 \pm 0.10$ | $0.45 \pm 0.04$ |

Figure 6 presents the validation of the proposed method for a sparse biological sample – human cheek cells. Fig. 6(a) shows the in-focus intensity measurement, while Figs. 6(b)-6(d) show the TIE, FPM and FPM+TIE reconstructions respectively. As can be observed, the FPM and FPM+TIE results reconstruct small cell features, which are invisible in TIE-only result. FPM-only result misses the low-frequency phase information, which is present in TIE and FPM+TIE results. Interestingly, the proposed FPM+TIE method managed to reconstruct large phase elements with phase value smaller than $-\pi$ leading to visible phase wraps, marked with yellow arrow in Fig. 6(d). This indicates that FPM+TIE is suitable for reconstructing optically thick elements, which, due to missing of low-frequency phase components, is often problematic in FPM technique [19].

Analyzing FPM only result, one may find it more informative from the morphological point of view, as the cells details are visible with higher contrast due to the absence of low-frequency information. On the other hand, FPM+TIE result may deliver more quantitative result of the measured phase values, useful e.g., for the cells dry mass calculation or the cell structure RI based segmentation. Nevertheless, the standalone FPM result can also be obtained from FPM+TIE data, allowing for hybrid high-frequency only and high-low frequency reconstructions. Alternatively, FPM+TIE phase may be numerically high-pass filtered to arbitrarily adjust the amount of low-frequency information in the result.

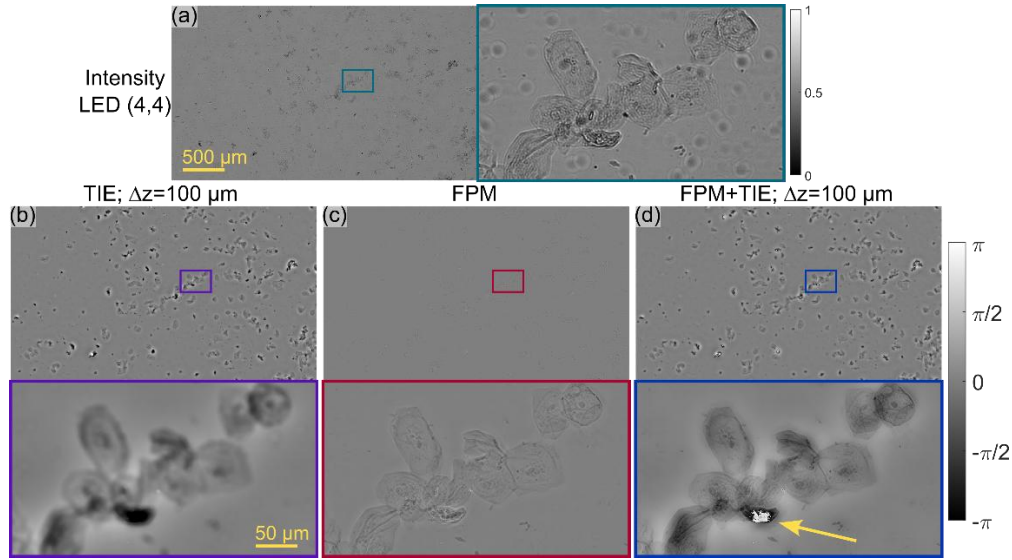

Fig. 6. (a) In-focus intensity image collected with central LED of the human cheek cells. (b)-(d) Phase reconstruction obtained with (b) TIE, (c) FPM and (d) FPM+TIE methods.

Figure 7 presents experimental validation of the proposed FPM+TIE approach for imaging cultured mouse neurons. Primary neuronal cultures were prepared following established protocols [46]: neurons were isolated from mouse brain tissue and cultured in vitro on thin glass coverslips using standard growth medium to support cell viability and adhesion. Once the cultures reached an appropriate stage for imaging, the coverslips with adherent neurons were mounted onto microscope slides and fixed using Fluoromount G.

The in-focus intensity image is shown in Fig. 7(a), while Figs. 7(b)–7(d) display phase reconstructions obtained using the TIE, FPM, and combined FPM+TIE methods, respectively. In both the TIE and FPM+TIE reconstructions, large optically thin membrane-like structures can be observed underlying the neurons. These likely correspond to densely clustered or confluent astrocytes, which are known to form sheet-like layers in primary cultures due to extensive outgrowth and adhesion.

Thanks to their improved sensitivity to low-frequency phase components, both TIE and FPM+TIE reveal these membranous features with significantly higher contrast than FPM alone, allowing clear differentiation between neurons located within such structures (e.g., green arrow in Fig. 7(d)) and those found in structure-free regions (e.g., yellow arrows). Compared to TIE, the FPM and FPM+TIE methods provide markedly enhanced resolution, enabling visualization of fine neuronal structures such as axons and dendrites (see zoomed-in insets in Fig. 7).

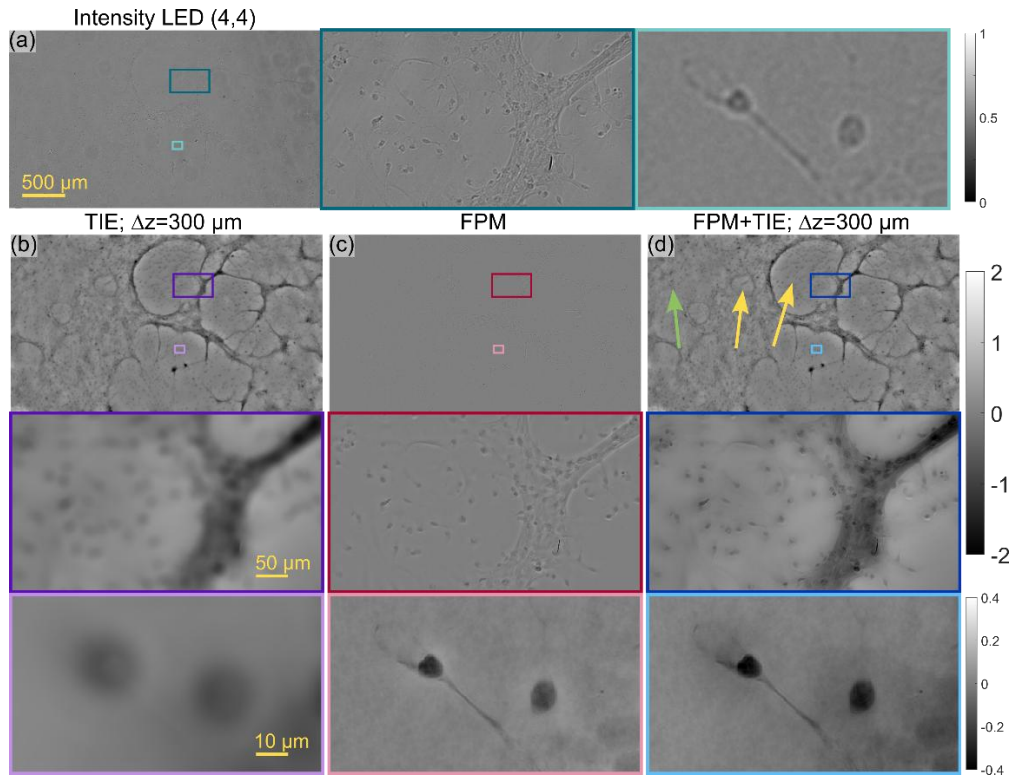

Fig. 7. (a) In-focus intensity image collected with central LED of the mouse neural cells. (b)-(d) Phase reconstruction obtained with (b) TIE, (c) FPM and (d) FPM+TIE methods.

Figure 8 shows the imaging results of a 40 μm thick slice of mouse brain tissue. The in-focus intensity image is presented in Fig. 8(a), with the corresponding TIE, FPM, and FPM+TIE phase reconstructions shown in Figs. 8(b)–8(d). In both FPM and FPM+TIE results, individual brain cells can be identified as phase-negative features (e.g., the one indicated with a green arrow in Fig. 8(d)), which likely correspond to structures with higher refractive index than the surrounding medium. These cells exhibit high contrast and are often surrounded by fine, lower contrast projections – morphological cues suggesting they may be pyramidal neurons located in the brain cortex.

Notably, the FPM+TIE reconstruction reveals an additional population of phase-positive cells (e.g., one marked with yellow arrow in Fig. 8(d)) that remain barely discernible in the FPM-only image. These features likely represent neurons located at a different depth, observed slightly out of focus, which nonetheless contribute meaningful low-frequency phase contrast.

This highlights the capability of the proposed hybrid method to recover low-frequency phase components that are otherwise lost in conventional FPM, thereby enhancing the visibility of both in-focus and out-of-focus cellular structures and enabling more reliable identification of individual cells within thick biological specimens.

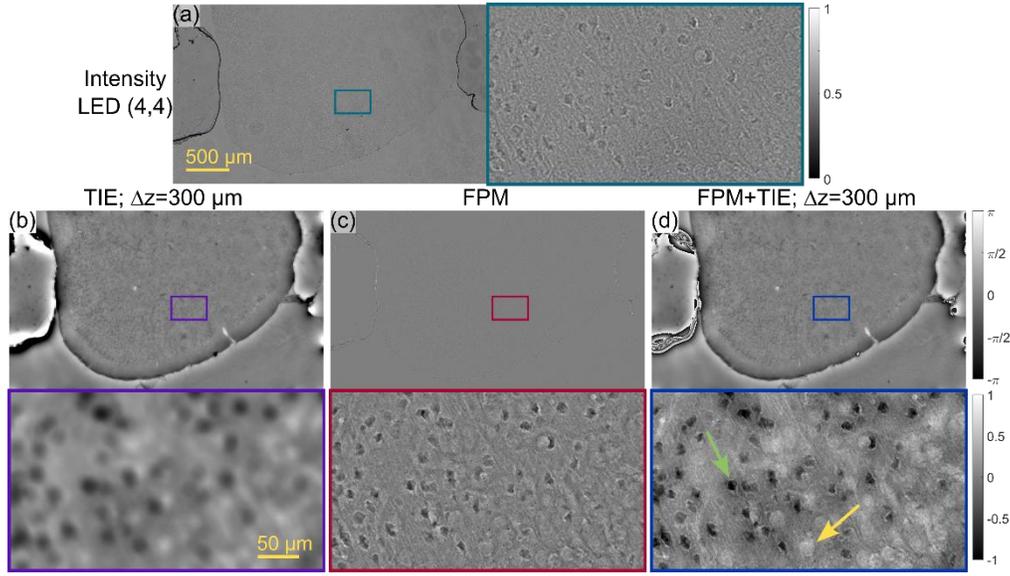

Fig. 8. (a) In-focus intensity image collected with central LED of the 40 μm thick mouse brain tissue slice. (b)-(d) Phase reconstruction obtained with (b) TIE, (c) FPM and (d) FPM+TIE methods.

## 5. Discussion and Conclusions

In this work, we presented a hybrid phase retrieval method that integrates transport of intensity equation and Fourier ptychographic microscopy to overcome the inherent limitations of each individual technique. The proposed approach combines the ability of TIE to capture low-frequency phase information with the high-resolution reconstruction capabilities of FPM, resulting in improved quantitative phase imaging performance across a wide range of spatial frequencies.

A key advantage of the method lies in its ability to recover the full phase spectrum even in the absence of matching NA conditions – a critical requirement in conventional FPM for accurate low-frequency phase recovery. While this condition is theoretically well-defined, it is often practically unattainable due to the spherical nature of the LED illumination used in typical FPM systems. As a result, only a small portion of the FOV typically satisfies the ideal illumination geometry, leading to inconsistent phase reconstructions. By incorporating TIE-based low-frequency estimation into the FPM reconstruction pipeline, the proposed method effectively bypasses this constraint, extending the applicability of FPM to realistic experimental setups.

Another important contribution of this work is the demonstration that our approach allows for the use of unusually large defocus distances in TIE, which significantly improves noise robustness. In conventional TIE, large defocus introduces strong blur and resolution loss, while small defocus is highly susceptible to low-frequency artifacts in the presence of noise. In our hybrid method, high spatial frequencies are recovered via FPM, which allows TIE to focus on accurate low-frequency reconstruction. As a result, large defocus distances – which are more resilient to noise – can be safely used, offering a practical advantage in noisy or photon-limited imaging conditions. Moreover, the method can be further enhanced through the use of multiple defocus planes, as is commonly done in advanced TIE-based techniques [23,38].

To enable accurate TIE reconstruction under large defocus conditions, we introduced a new global affine transformation model that compensates for the magnification mismatch between the intensity images acquired at different defocus planes. This effect arises due to the spherical nature of the illumination and becomes especially pronounced for large Δz values. Although this correction is not required for the FPM reconstruction itself, it is essential for aligning the defocused images used in TIE and for ensuring consistent scaling before merging the TIE and FPM results.

Importantly, our results also indicate that the method holds promise for imaging optically thick samples, where conventional FPM often fails due to the absence of low-frequency phase components. Without these components, phase discontinuities resulting from steep refractive index gradients cannot be properly unwrapped, leading to severe artifacts or incorrect reconstructions. In our hybrid method, the inclusion of TIE-based low-frequency information enables a more faithful reconstruction of such features, as illustrated in the imaging of human cheek cells where wrapped phase features were observed. While this observation was not the primary focus of this study and requires further investigation, it highlights an important potential application of the proposed technique.

Experimental validation on neuron cultures and mouse brain slices confirmed the practical benefits of the method in biological imaging. In both cases, the hybrid FPM+TIE reconstruction enabled clear identification of features that were either weakly visible or entirely lost in standalone TIE (due to low resolution) or FPM (due to lack of low phase spatial frequencies) results. This illustrates the method's potential to facilitate more accurate segmentation, morphological analysis, and label-free visualization of transparent biological structures.

In conclusion, the FPM+TIE hybrid method provides a robust, flexible, and experimentally feasible framework for quantitative phase imaging over large fields of view and across a broad spectrum of spatial frequencies. It addresses key limitations of FPM in practical settings, improves phase reconstruction fidelity in noisy conditions, and opens the door to new applications in imaging of optically thick samples. Future work may include a systematic analysis of defocus optimization, extension to multiple-plane TIE integration, and further exploration of its capabilities in biomedical imaging.

**Funding.**

Research was funded by the Warsaw University of Technology within the Excellence Initiative: Research University (IDUB) programme (YOUNG PW 504/04496/1143/45.010008) and project no. WPC3/2022/47/INTENCITY/2024 funded by the National Centre for Research and Development (NCBR) under the 3rd Polish-Chinese/Chinese-Polish Joint Research Call (2022). The research was conducted on devices cofounded by the Warsaw University of Technology within the Excellence Initiative: Research University (IDUB) programme. National Natural Science Foundation of China (62361136588).

**Data availability.**

Data underlying the results presented in this paper are not publicly available at this time but may be obtained from the authors upon reasonable request.